\documentclass{isprs} 
\usepackage{subfigure}
\usepackage{setspace}
\usepackage{geometry} 
\usepackage{epstopdf}
\usepackage[labelsep=period]{caption}  
\usepackage[british]{babel} 
\usepackage[hang]{footmisc}
\usepackage{url}
\usepackage{caption}
\begin{document}

\title{Transfer of Manure from Livestock Farms to Crop Fields as Fertilizer using an Ant Inspired Approach}


\author{
 A. Kamilaris\textsuperscript{1,2,}\thanks{Corresponding author}, A. Engelbrecht\textsuperscript{3}, A. Pitsillides\textsuperscript{4}, Francesc X. Prenafeta-Boldú\textsuperscript{5}}

\address{
	\textsuperscript{1 }Research Centre on Interactive Media, Smart Systems and Emerging Technologies (RISE),
Nicosia, Cyprus - a.kamilaris@rise.org.cy\\
	\textsuperscript{2 }Department of Computer Science, University of Twente, The Netherlands - a.kamilaris@utwente.nl\\
	\textsuperscript{3 }Department of Industrial Engineering, University Of Stellenbosch, South Africa - engel@sun.ac.za\\
	\textsuperscript{4 }Department of Computer Science, University of Cyprus, Nicosia, Cyprus - andreas.pitsillides@ucy.ac.cy\\
		\textsuperscript{5 }Institute of Agriculture and Food Research and Technology (IRTA), Barcelona, Spain - francesc.prenafeta@irta.cat}


\icwg{}   

\abstract{
Intensive livestock production might have a negative environmental impact, by producing large amounts of animal excrements, which, if not properly managed, can contaminate nearby water bodies with nutrient excess.
However, if animal manure is exported to distant crop fields, to be used as organic fertilizer, pollution can be mitigated.
It is a single-objective optimization problem, in regards to finding the best solution for the logistics process of satisfying nutrient crops needs by means of livestock manure. 
This paper proposes a dynamic approach to solve the problem, based on a decentralized 
nature-inspired cooperative technique, inspired by the foraging behavior of ants (AIA). 
Results provide important insights for policy-makers over the potential of using animal manure as fertilizer for crop fields, while AIA solves the problem effectively, in a fair way to the farmers and well balanced in terms of average transportation distances that need to be covered by each livestock farmer.
Our work constitutes the first application of a decentralized AIA to this interesting real-world problem, in a domain where swarm intelligence methods are still under-exploited. 
}


\maketitle

 
\sloppy


\section{INTRODUCTION}
\label{intro}
The central role of the agricultural sector is to provide adequate and high-quality food to an increasing human population,
which is expected to be increased by more than 30\% by 2050 \cite{UNFood}. This means that a significant increase in food production must be achieved.
Because of its importance and relevance, agriculture is a major focus of policy agendas worldwide.
Agriculture is considered as an important contributor to the deterioration of soil, water contamination, as well as air pollution and climate change \cite{bruinsma2003world}, \cite{vu2007survey}.
Intensive agriculture has been linked to excessive accumulation of soil contaminants \cite{teira2003method},
and significant groundwater pollution with nitrates \cite{stoate2009ecological}, \cite{garnier1998integrated}.

\begin{figure}[ht!]
\begin{center}
\includegraphics[width=1.0\columnwidth]{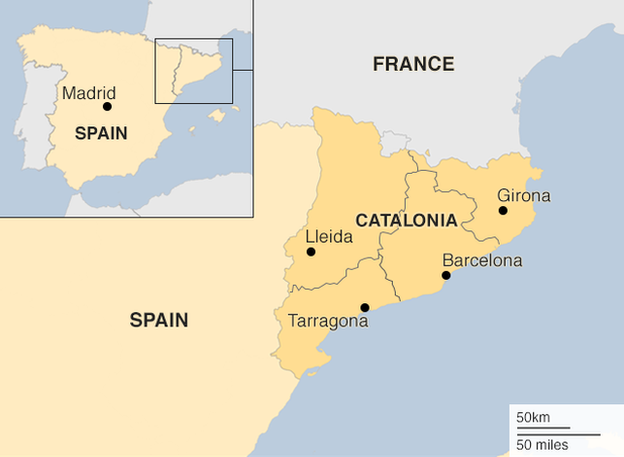}
\caption{Geographical map of Catalonia, Spain.}
\label{fig:Catalonia}
\end{center}
\end{figure}

In particular, livestock farming could have severe negative environmental effects \cite{heinrich2014meat}.
Livestock farms produce large amounts of animal manure, which, if not properly managed, can contaminate nearby underground and aboveground water bodies \cite{cheng2007non}, \cite{infascelli2010environmental}, \cite{vu2007survey}.
The autonomous community of Catalonia, located at the north-east part of Spain near the borders with France (see Figure \ref{fig:Catalonia}), is facing this challenge, as livestock farming (mainly swine) has 
contributed to the pollution of the physical environment of the area during the last decades \cite{Kamilaris2017AgriBigCat}, \cite{Kamilaris2018CNNAgri12}.
The high density of livestock in some areas, linked to insufficient accessible arable land, has resulted in severe groundwater pollution with nitrates \cite{directive1991council}. 
Catalonia is one of the European regions with the highest livestock density, according to the agricultural statistics for 2016, provided by the Ministry of Agriculture, Government of Catalonia,
with reported numbers of around 7M pigs, 1M cattle and 32M poultry in a geographical area of 32,108 km{$^2$}.

If handled and distributed properly, manure can be applied as organic fertilizer in crop fields that produce different types of fruits and cereals, nuts and vegetables. In this way, the potential contamination of soil and water created by animal manure could be mitigated \cite{he1998preliminary}, \cite{teira2003method}, \cite{paudel2009geographic},
while a positive effect on soil acidity and nutrient availability is possible \cite{whalen2000cattle}.
Hence, if the animal manure is efficiently exported at specific seasons of the year to nearby or distant crop fields, manure can eventually become a valuable resource rather than waste \cite{keplinger2006economics}, \cite{teenstra2014global}, \cite{oenema2007nutrient}.
To achieve this aim in an optimal manner, the costs of transporting large quantities of manure must be taken into account as a limiting factor in the process of nutrients' transfer from livestock farms to agricultural fields.

This paper proposes a nature-inspired method to solve the issue of transporting manure from livestock farms to crop fields,
to be used as fertilizer in the territory of Catalonia. This method is a decentralized approach, motivated by the synergistic behaviour of ants at the task of depositing pheromone near food sources, in order to attract more ants to follow their trajectory. This task is foraging, which is achieved by following pheromone trails, and depositing more pheromone on trails during their traversal.
This task creates in a synergistic way promising paths in terms of discovering food \cite{bonabeau1999swarm}, \cite{garnier2007biological}, \cite{paredes2017milk}. Intuitively, it can be applied in the context for discovering crop farms in need of fertilizer, similar to the way it has been applied in the past to solve a milk collection problem \cite{paredes2017milk}.

Our contribution in this paper is two-fold: on the one hand, we have solved the problem of transferring animal manure in a decentralized way, addressing some limitations of related work (see Section \ref{relWOrk}).
On the other hand, we have proposed and developed a nature-inspired technique for a domain (i.e. smart agriculture)
where swarm intelligence methods are still under-exploited, although there is a growing research interest from a computational science perspective \cite{KamilarisSept2018NatComp}.
To our knowledge, it is the first attempt to use an ant-inspired algorithm (AIA) for this particular and challenging real-world problem.

The rest of the paper is organized as follows:
Section \ref{relWOrk} describes related work on manure management based on geospatial analysis and on ant-inspired applications in agriculture,
while Section \ref{Methodology} presents our methodology employing an ant-inspired modelling approach (AIA).
Section \ref{Results} analyzes the overall findings after applying the proposed method in the Catalonian context,
and Section \ref{Discussion} comments on the perspectives of this research suggesting future work. Finally, Section \ref{Conclusion} concludes the paper.         

\section{RELATED WORK}
\label{relWOrk}
Related work involves two main research areas: manure management based on geospatial analysis, facilitated by Geographical Information Systems (GIS) \cite{Kamilaris2018CNNAgri},
as well as applications of ant-inspired techniques in agriculture, facilitated by ant colony optimization (ACO) \cite{dorigo1996ant}, \cite{dorigo1997ant}. Less relevant work is about network flow solutions applied to other agricultural problems, such as dealing with transportation of live animals to slaughterhouses \cite{oppen2008tabu}, the routing of vehicles for optimized livestock feed distribution \cite{kandiller2017multi} or for biomass transportation \cite{gracia2014application} etc.
Related work in the two main research areas mentioned above is presented below.

\subsection{Anti-Inspired Techniques in Agriculture}
Not much research has been done in applying ant-inspired techniques in agriculture.
Some research has been performed on the application of ACO in various agricultural problems.
ACO generally works by searching for optimal paths in a graph, based on the behaviour of ants seeking a path between their colony and sources of food.
Paredes et al. \cite{paredes2017milk} applied ACO to solve the milk blending problem with collection points, determining where the collection points should be located and which milk producers would be allocated to them for delivery.
Optimal land allocation was investigated in \cite{liu2012multi}, where the ants represented candidate solutions for different types of land use allocation.
Li et al. (2010) used an ACO algorithm for feature selection in a weed recognition problem \cite{li2010shape}.
Optimization of field coverage plans for harvesting operations was performed by means of ACO \cite{bakhtiari2013operations}. Finally, ACO was used for
feature selection and classification of hyperspectral remote sensing images \cite{zhou2009feature}, an operation highly relevant to agriculture.

\subsection{Transport of Manure for Nutrient Use}
The idea of transporting surplus manure beyond individual farms for nutrient utilization was proposed in \cite{he1998preliminary},
focusing on animal manure distribution in Michigan.
\cite{teira2003method} proposed a methodology to apply manure at a regional and municipal scale in an agronomically correct way,
i.e. by balancing manure distribution to certain crops, based on territorial nitrogen needs and also based on predictions of future needs and availability considering changes in land use.
ValorE \cite{acutis2014valore} is a GIS-based decision support system for livestock manure management,
with a small case study performed at a municipality level in the Lombardy region, northern Italy,
indicating the feasibility of manure transfer.

Other researchers proposed approaches to select sites for safe application of animal manure as fertilizer to agricultural land \cite{van1992computer}, \cite{basnet2001selecting}.
Site suitability maps have been created using a GIS-based model in the Netherlands and in Queensland, Australia respectively. 
In \cite{van1992computer}, 40\% to 60\% of Dutch rural land was found suitable for slurry injection, while 16\% of the area under study was found suitable for
animal manure application in \cite{basnet2001selecting}.
A minimum cost spatial GIS-based model for the transportation of dairy manure was proposed in \cite{paudel2009geographic}.
The model incorporated land use types, locations of dairy farms and farmlands, road networks, and distances
from each dairy farm to receiving farmlands, to identify dairy manure transportation routes that minimize costs relative to environmental and economic constraints.

The aforementioned related work has adopted the following assumptions/limitations:
\begin{enumerate}

 \item aggregating geographical areas at county-level \cite{he1998preliminary};
 \item selecting generally suitable sites (i.e. crop and pasture areas) to apply animal excrements \cite{van1992computer}, \cite{basnet2001selecting};
 \item not considering transportation distances between livestock and crop farms \cite{he1998preliminary}, \cite{teira2003method};
 \item not calculating the particular needs of crop fields in nitrogen that depend on the land area and the type of the crop \cite{basnet2001selecting}, \cite{paudel2009geographic};
 \item not including actual costs involved with the proposed solution \cite{he1998preliminary}, \cite{paudel2009geographic}, \cite{teira2003method}, \cite{basnet2001selecting}; 
 \item not finding a balanced, fair solution that minimizes the average distance that needs to be covered by the livestock farmers (all aforementioned papers); and
 \item approximating the problem by means of (only) centralized, static strategies (all aforementioned papers).
 \end{enumerate}
The work of this paper tries to eliminate or partially address these assumptions.

\section{PROBLEM MODELLING AND METHODS}
\label{Methodology}
The overall goal is to solve the problem of how to find an optimal and economic way to distribute animal manure in order to fulfil agricultural fertilization needs.
The purpose of this section is to describe how the problem was modelled using the area of Catalonia as a case study.

\subsection{Problem Modelling}
\label{problemmodel}
\begin{figure*}[ht!]
\begin{center}
\begin{tabular}{l}
	\begin{minipage}{\linewidth}  
		\begin{minipage}{0.50\linewidth}
		\includegraphics[width=\linewidth]
			{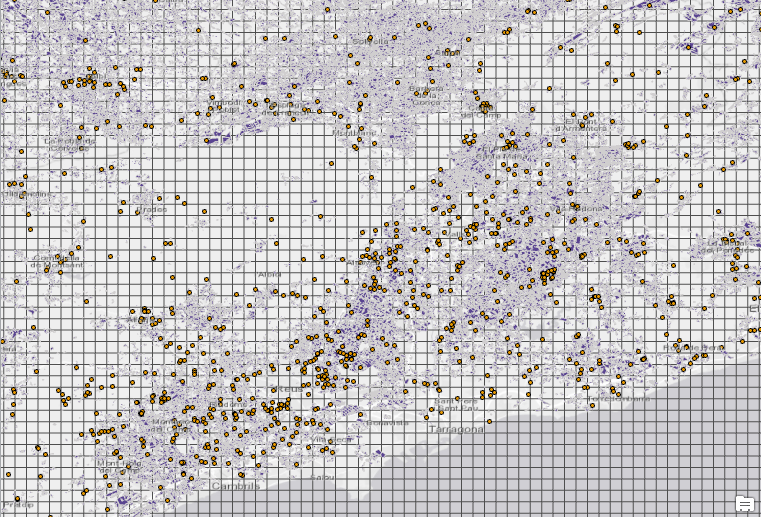}	
		\end{minipage}
		\begin{minipage}{0.49\linewidth}
		\includegraphics[width=\linewidth]
			{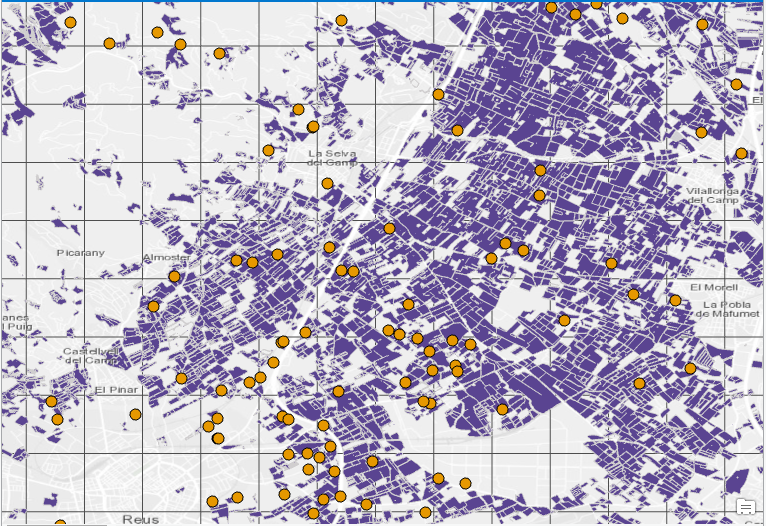}
		\end{minipage}
	\end{minipage}
\end{tabular}
\end{center}
\vspace{-0.3cm}
\caption{Division of the area of Catalonia in cells of 1 square kilometre each (left). Example grid cells in a dense agricultural area of the region (right). This is a zoom of the map on the left. Livestock farms are shown in brown, crop fields in blue.}
\label{fig:CataloniaModel}
\end{figure*}

To simplify the problem, the geographical area of Catalonia has been divided into a two-dimensional grid, as shown in Figure \ref{fig:CataloniaModel} (left).
In this way, the distances between livestock farms (i.e. original grid cell) and crop fields (e.g. destination grid cell) are easier to compute, considering straight-line grid
cell Manhattan distance as the metric to use (and not actual real distance through the existing transportation network). The centre of the crop field is used for calculations. An approximation to real-world distances is attempted in Section \ref{objFunctionDescription}.

Each crop and livestock farm has been assigned to the grid cell where the farm is physically located, as depicted in Figure \ref{fig:CataloniaModel} (right).
Brown small circles represent livestock farms while blue areas depict crop fields.
Details about livestock farms (i.e. animal types and census, location etc.) have been provided by the Ministry of Agriculture of Catalonia for the year 2016, after signing a confidentiality agreement.
Details about crop fields (i.e. crop type, hectares, irrigation method, location etc.) have been downloaded from the website of the Ministry\footnote{Ministry of Agriculture of Catalonia. \url{http://agricultura.gencat.cat/ca/serveis/cartografia-sig/aplicatius-tematics-geoinformacio/sigpac/}},
for the year 2015.
For every livestock farm, the yearly amount of manure produced and its equivalent in nitrogen as fertilizer have been
calculated, depending on the type and number of animals on the farm, based on the IPCC guidelines (TIER1) \cite{IPCC2006} and the work in \cite{borhan2012greenhouse}. 
Similarly, for every crop field, the yearly needs in nitrogen have been computed, depending on the crop type and total hectares of land,
according to \cite{RuralCatdossier}.

The estimated total fertilizer needs of crop fields (i.e. 81,960 tons of nitrogen) were lower than the availability of nitrogen from animal manure (i.e. 116,746 tons of nitrogen).
This means that the produced amount of manure/nitrogen from livestock agriculture has the potential to completely satisfy the total needs of crop farms. This would be particularly important in areas corresponding to the vulnerable zones
defined by the nitrogen EU directive\footnote{The Nitrates Directive of the European Commission. \url{http://ec.europa.eu/environment/water/water-nitrates/index_en.html}}.

Summing up, the total area of Catalonia has been divided into 74,970 grid cells, each representing a $1 \times 1$ square kilometre of physical land. 
Every cell has a unique ID and $(x,y)$ coordinates, ranging between $[1,315]$ for the $x$ coordinate and $[1,238]$ for the $y$ coordinate.
For each grid cell, we are aware of the crop and livestock farms located inside that cell, the manure/nitrogen production (i.e. from the livestock farms) and the needs in nitrogen (i.e. of the crop fields).

\subsection{Objective Function}
\label{objFunctionDescription}
The problem under study is a single-objective problem, with the overall goal of optimizing the logistics process of satisfying nutrient crops needs by means of livestock waste. This goal has the following (conflicting) sub-objectives:
\begin{enumerate}
 \item The total fertilizer needs at the crop fields have to be satisfied as much as possible.
 \item The total aggregated travel distance covered from the livestock farms to the crop fields, in order to deposit the manure/fertilizer, needs to be as short as possible.
\end{enumerate}

These two sub-objectives can be reformulated as a single one by combining them linearly, assuming the following:
\begin{itemize}
 \item The price of fuel in Catalonia, Spain is 1.27 Euro per liter\footnote{GlobalPetrolPrices. \url{http://es.globalpetrolprices.com/Spain/gasoline_prices/} (for May 2019)}.
 \item The fuel consumption of tanks is 20.3 liters per 100 kilometres\footnote{Natural Resources Canada. \url{http://www.nrcan.gc.ca/energy/efficiency/transportation/cars-light-trucks/buying/16745}}.
 This is equivalent to 0.203 liters per kilometre.
 \item Based on the price of fuel in Spain, as given above, the transportation cost per kilometre is 0.257 Euro.
 \item Based on the local monthly average prices for fertilizers in Catalonia\footnote{Ministry of Agriculture of Catalonia. \url{http://agricultura.gencat.cat/ca/departament/dar_estadistiques_observatoris} (ammonium sulphate in May 2019)},
 the value of nitrogen is 22,50 Euro per 100 kilograms or 0.225 Euro per kilogram.
\end{itemize}

Based on the aforementioned assumptions, the general objective is defined as:
\begin{equation}
\label{combinedObjective}
 GO = (NT \times 0.225 \times l) - (TD \times 0.257 \times g)
\end{equation}

where $NT$ is the total nitrogen transferred in kilograms, and $TD$ is the total distance in kilometres
covered to transport manure, from the livestock to the crop farms. The parameter $l$ aims to capture the nutrient losses of manure during its storage time,
i.e. the time when the manure is stored at the livestock farm until it is transferred to the crop field. Depending on animal type and storage method, nutrient losses vary.
We selected a value of $l=0.60$, which is the average percentage of nitrogen remaining availability in manure according to the animal census of Catalonia,
at an expected storage time of up to three months as solid or liquid manure \cite{rotz2004management}.
Moreover, the parameter $g$ is an approximation of real-world distance, based on the Manhattan distance used in the calculations of travel distance from the livestock to crop farms. 
$g$ weights the calculated Manhattan distance by a factor of $g = 1.30$, a value appropriate for semi-rural landscapes \cite{wenzel2017comparing}. 

The objective $GO$ is assumed to be in Euro, representing a simplified cost/benefit relationship of the manure transfer problem, i.e. benefit of selling nitrogen to the crop fields and cost of transport needed in order to transfer the nitrogen.
The overall goal is to maximize $GO$, whose value can be translated to gains or losses of each solution of the problem.
$GO$ can take negative values if some solution had produced a loss.

Moreover, there is a hard constraint set by the Ministry of Agriculture, demanding that the maximum distance travelled for manure deposit is $50$ kilometres. The reasoning behind this is that (otherwise) the travel time required for the transfer would have become significant and should have somehow become included in the calculations. Finally, the Ministry asked to try to maintain the average travel distance (and standard deviation) from every livestock farm to the crop fields as small as possible, i.e. to keep the proposed solution \textit{well-balanced and fair} for all livestock farms.

\subsection{Ant-Inspired Algorithm}
\label{AIA}
In general, the synergistic pheromone laying behavior of ants when discovering food sources
is used as a form of indirect communication, in order to influence the movement of other ants \cite{bonabeau1999swarm}, \cite{garnier2007biological}.
Pheromone laying was modelled (among others) in the Ant System \cite{dorigo1996ant}, \cite{dorigo1997ant}, a probabilistic population technique
for combinatorial optimization problems where the search space can be represented by a graph.
The technique exploits the behaviour of ants following links on the graph, constructing paths between their colony and sources of food, to incrementally discover optimal paths, which would form the solution.



In the particular context of the manure transport problem, the foraging behavior of ants has been adapted  to the problem under study.
The modelling of the problem according to ant foraging is as follows:
\begin{enumerate}
 \item Every livestock farm simulates an ant.
 \item Every crop field is considered as a potential source of food, analogous to its needs in nitrogen.
 \item Ants perform local pheromone updates (to the grid cell where they are currently located while moving around) proportional to the amount of food available (i.e. nitrogen needs) in their grid-based neighbourhood of Manhattan distance (radius) $n$.
 \item Pheromone at each grid cell is updated by pheromone deposits. 
  At the beginning, the pheromone amount at each grid cell is initialized proportionally to the initial needs in nitrogen by the crop fields physically located inside the grid cell.
 \item Each ant chooses the next link of its path based on information provided by other ants, in the form of pheromone deposits at every grid cell.
  \item The pheromone value at each grid cell, created by the ants which have resided  at the cell at some particular iteration of the algorithm, increases when one or more ants reside at the cell at some point, depositing pheromone, but also evaporates with time.  
 \item Whenever an ant discovers a  crop field with nitrogen needs at its current position (i.e. some grid cell), a transfer of nitrogen is performed from the livestock farm represented by the ant, to the crop field located at that grid cell. In this case, the need for nitrogen at that particular grid cell is reduced accordingly. The manure transaction is recorded by the system as part of the final solution.
 \item If the ant still carries some manure/nitrogen, then it continues to move in the grid up to a maximum Manhattan cell-distance of $m=50$ from its initial position.
\end{enumerate}

Each ant (i.e. livestock farm) selects its next position from its current grid position successively
and pseudo-randomly, where the probability of next move depends on the pheromone amounts at the neighbouring grid cells.
At each iteration of the algorithm, each ant is allowed to move at a Manhattan distance of maximum one neighbouring grid cell.
Each ant examines the availability of nitrogen needs by crop fields in its neighbourhood,
and drops pheromone at its current grid cell, proportional to the local needs in nitrogen, in order to inform other ants of the demand in manure at nearby crop fields.

The amount of pheromone laid by each ant is calculated based on the amount of existing nitrogen needs at each neighbouring cell within radius $n$,
and the Manhattan distance between the ant's current location and the neighbouring grid cells.
The Manhattan distance calculated is used to penalize neighbours at larger distances, reducing their \textit{contribution} to the pheromone deposits.
The amount of pheromone $\tau_{xy}$, laid by each ant located at grid cell $(x,y)$ at every iteration $t$ of the algorithm, is calculated using:

\begin{equation}
\label{pheromonecreation}
 \tau_{xy}(t) \, = \, \tau^\prime_{xy}(t-1) + \sum_{i=x-n}^{x+n} \sum_{j=y-n}^{y+n} NN_{ij} \times \frac{1}{ d_{ijxy}}
\end{equation}

where $\tau_{xy}(t-1)$ is the previous concentration of pheromone at grid cell $(x,y)$,
$NN_{ij}$ represents the food (i.e. needs in nitrogen of the crop field in kilograms) located at grid cell $(i,j)$,
and $d$ is the Manhattan distance between the ant (i.e. livestock farm) and the food (i.e. crop field). 
The parameter $n$ defines which neighbors at the grid structure would be involved in the calculations of pheromone (i.e. neighbours up to $n$-cell distance).

The probability $p_{kl}$ of an ant to move from grid cell $(x,y)$ to $(k,l)$, is calculated as:
\begin{equation}
\label{antmove}
 p_{kl} \, = \, \frac{\tau_{kl}} {\sum_{i=x-1}^{x+1}\sum_{j=y-1}^{y+1} \tau_{ij} }
\end{equation}

Note that paths with a higher pheromone concentration have higher probability of selection.

At each iteration $t$ of the algorithm, the pheromone concentration $\tau_{xy}(t)$ at every grid cell $(x,y)$ decays/evaporates to promote exploration:
\begin{equation}
\label{pheromoneevap}
 \tau_{xy}(t) \, = \, (1-\varrho) \times \tau^\prime_{xy}(t-1)
\end{equation}
where $\varrho$ is the percentage of \textit{pheromone evaporation}.

\begin{figure}[ht!]
\centering
\includegraphics[width=1.0\columnwidth]{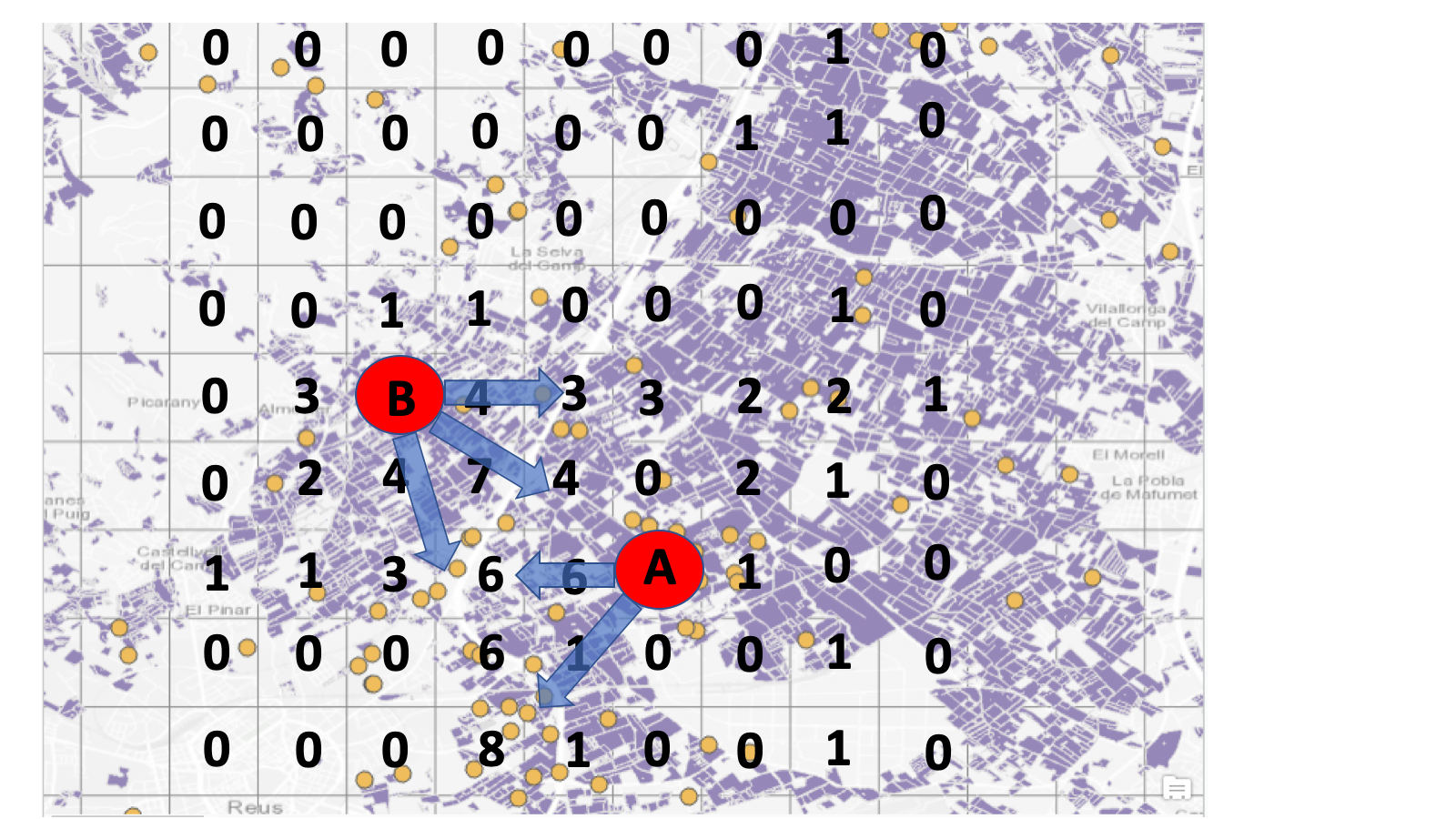}
\caption{Intuitive behavior of the AIA algorithm.}
\label{fig:behaviorAIA}
\end{figure}

Intuitively, the behavior of the AIA algorithm is described in Figure \ref{fig:behaviorAIA}, using an example of two ants (i.e. livestock farms), named A and B, located nearby in the grid structure of the map of Catalonia. Ant A started from a position close to B, depositing pheromones at the grid cells it has travelled, along its probabilistic transition to its current position, creating a pheromone trail. B follows this trail by sensing the pheromones. Multiple promising paths are available to B, as shown by the blue arrows. B will eventually choose one to follow probabilistically. At the same time, if A smells more pheromones nearby, in comparison to its current position, then it continues its exploration probabilistically as well, releasing new pheromones. This procedure continues iteratively and synergistically, until the whole simulation reaches an equilibrium or until there are no better paths to explore.

The ant-inspired algorithm introduces the control parameters $n$ and $\varrho$.
Additionally, two more parameters involved in our model are the \textit{maximum cell-distance} $m$ and the \textit{maximum number of iterations}. The former refers to the maximum Manhattan distance
between livestock and crop farms, where nitrogen transfer could be allowed, while the latter defines the maximum number of iterations until the algorithm stops.
The algorithm could stop earlier if no more transfers occur (i.e. all needs are satisfied or no more manure is available).
All parameters involved in the model are listed in Table \ref{tab:ParametersAIS} while the selection of values for these parameters is discussed in Section \ref{AISparameterSetting}.

\begin{table*}[h]
\centering
\label{tab:ParametersAIS}       
\begin{tabular}{| p{3.6cm} | p{7.0cm} | p{5.0cm} |} \hline
\bf{Parameter Name} & \bf{Description} & \bf{Value(s)} \\ \hline
Pheromone evaporation, $\varrho$ & The decay of pheromone deposited by the ants, at each iteration of the algorithm. & 0-100\% \\

Neighborhood radius, $n$ & The maximum Manhattan distance, at which neighbouring cells will contribute in calculating pheromone that would be released by the ant. All the cells up to a cell distance $n$ participate in the calculations. & 1-50 grid cells (values up to 65 have been allowed only for testing purposes)\\

Minimum nitrogen & The minimum amount of nitrogen in kilograms for a transfer to occur, yielding a positive value of the objective $GO$. & 1-150 Kilos, depending on the Manhattan distance between farms. \\

Maximum cell-distance, $m$ & The maximum Manhattan distance over which transport of animal manure/nitrogen is allowed. & 50 grid cells (values up to 60 have been allowed only for testing purposes)\\

Maximum iterations & The maximum number of iterations of the AIA algorithm. & 3,000 \\
\hline
\end{tabular}
\caption{Control parameters for the AIA algorithm.}
\end{table*}

\section{ANALYSIS AND RESULTS}
\label{Results}
This section first explains the reasoning towards the tuning of the control parameters of the AIA. Then, it presents the findings obtained by solving the problem of manure transport optimization using AIA.

\subsection{Parameter Tuning}
\label{AISparameterSetting}
The different values of the control parameters of the AIA have been listed in Table \ref{tab:ParametersAIS}.
These parameters under study are the neighbourhood distance $n$ and the pheromone evaporation coefficient $\varrho$. The former takes values in the range $[0,65]$ (ignoring here for reasons of comparison the hard constraint of 50 kilometres), while the latter takes values in the range $[0,100]$.

\begin{figure}[ht!]
\centering
\includegraphics[width=1.0\columnwidth]{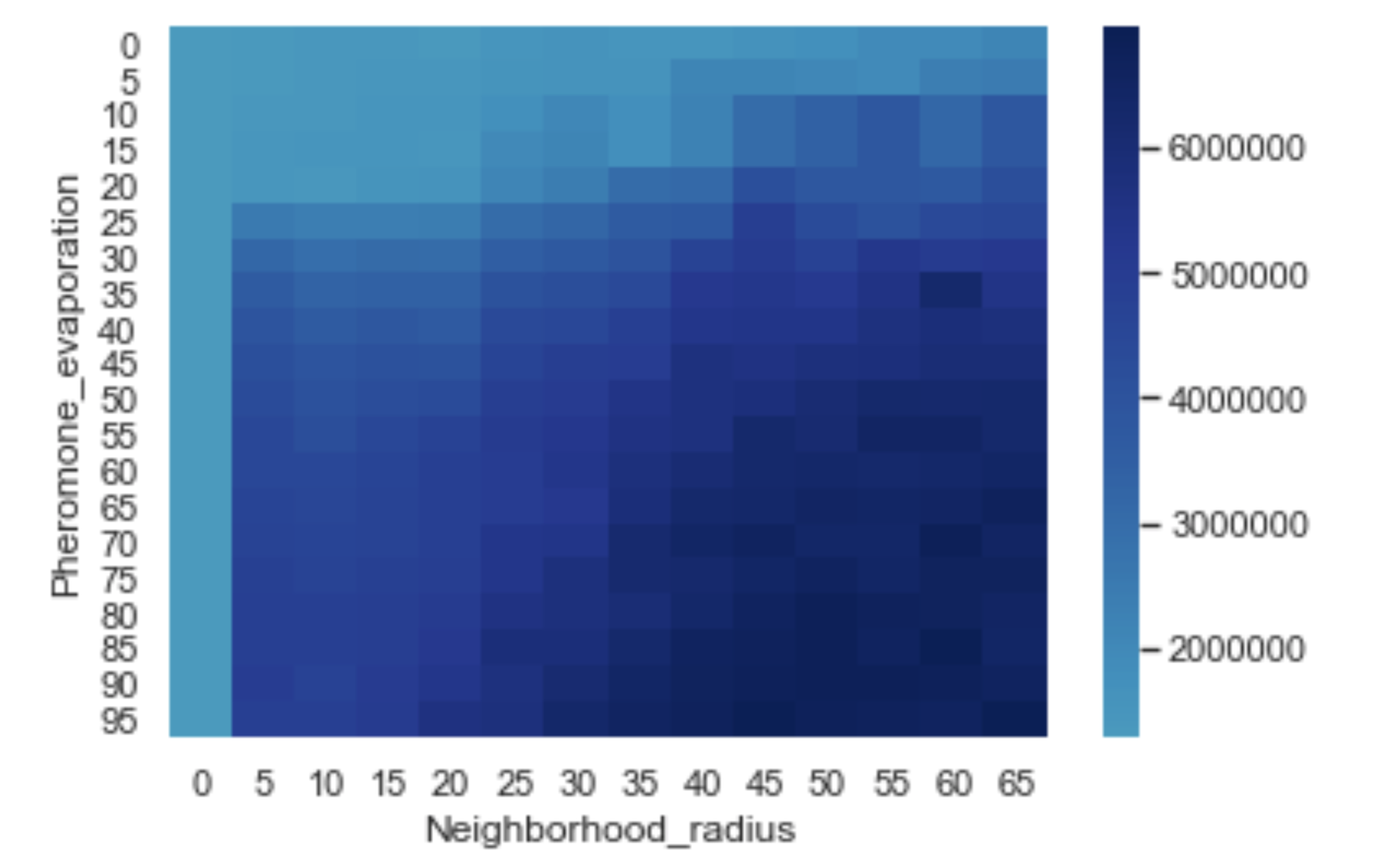}
\caption{Impact of pheromone evaporation $\varrho$ and neighbourhood radius $n$ on the objective $GO$.}
\label{fig:paramsAIA}
\end{figure}

Figure \ref{fig:paramsAIA} depicts the different values of the objective $GO$, at different values of distance $n$ and percentages of $\varrho$. Note that, because the AIA algorithm is stochastic, the results presented below have been averaged over 10 independent runs of the algorithm, with different value pairs of control parameters. The maximum value was recorded for each value pair. Differences between experiments with the same value pairs were very small.

Based on the results presented in Figure \ref{fig:paramsAIA}, a value of pheromone evaporation $\varrho=85\%$ 
and a neighbourhood radius $n=50$ cells-distance
were selected, because this combination of  values maximized the $GO$ (see Section \ref{SolFin}). We note that values of $n$ larger than the hard constraint of 50 kilometres did not improve $GO$, and have been included for comparisons. We also note that values of $\varrho \in [85,95]$ and $n \in [50,65]$ resulted in very small differences in the $GO$ value.

\subsection{Solution and findings}
\label{SolFin}
Table \ref{tab:summary} summarizes the results of the experiment for a duration of one complete year. Around 550 thousand kilometres are required to transfer 51 tons of nitrogen in total. These values consider the aggregated transfers performed by all the livestock farmers of 
The 3rd and 4th rows of the table denote the average total Manhattan distance that needs to be travelled by each livestock farmer and the standard deviation respectively, in order to perform transfer(s) of animal manure. This average distance is 57 for the AIA method (with std. deviation of 25). This relates to the requirement stated in Section \ref{objFunctionDescription}, i.e. the proposed solution must be well-balanced and fair for all livestock farms.
The last row shows the running time of AIA in minutes, on a laptop machine (2,8 GHz Intel Core i7, 6 GB 2133 MHz LPDDR3 RAM).

\begin{table}[ht!]
\label{tab:summary}       
\centering
\begin{tabular}{| p{5.5cm} | c |} \hline
\bf{Objective} & \bf{AIA} \\ \hline
Nitrogen transferred (tons)  &  51.124  \\

Transportation (Manhattan distance)  & 549,829  \\

Objective $GO$ (Euro)  &  6.718,069   \\

Average transportation distance of each livestock farm (Manhattan distance)   & 57 \\

Standard deviation of the average transportation distance of each livestock farm (Manhattan distance)  & 25 \\ 

Run time (minutes)  &  38   \\
\hline
\end{tabular}
\caption{Summarized values of the experiment performed, considering one complete year.}
\end{table}

\begin{figure*}[ht!]
\centering
\includegraphics[width=1.0\linewidth]{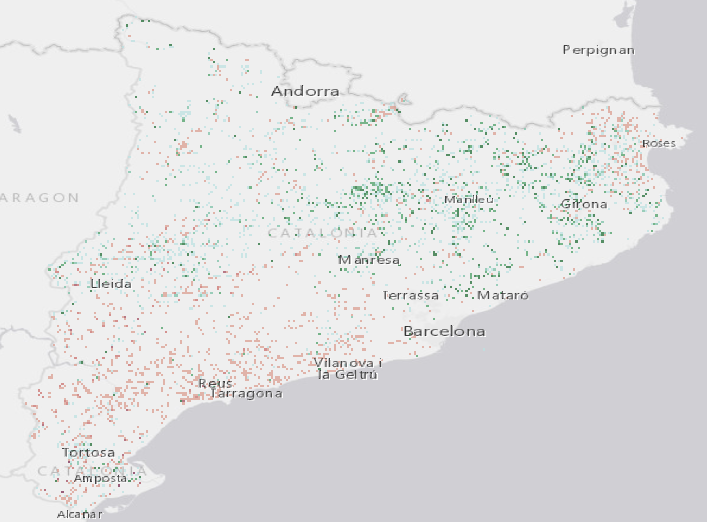}
\caption{The map of Catalonia after the AIA has been applied, showing remaining needs in manure (orange color) and remaining availability of manure (green color). The color intensity indicates different needs or availability of manure. For example, darker colours of green and orange correspond to larger availability or needs of manure at some farm. Please note that this map depicts only manure availability and needs of farms after the application of AIA. This means that livestock farms whose manure availability is zero and/or crop farms whose needs in manure as fertilizer are zero, do not appear on the map.}
\label{fig:AIAappliedCat}
\vspace{-0.1cm}
\end{figure*}

Figure \ref{fig:AIAappliedCat} illustrates how the application of AIA in the area of Catalonia affects availability (i.e. green colour) and needs (i.e. orange colour) of manure/nitrogen. We can observe that the algorithm creates separate regions of green- and orange-coloured spots (i.e. livestock and crop farms respectively). The distance between spots of different colour is either larger than 50 kilometres, or there is not enough manure available for the transaction to be gainful, i.e. give positive values to the $GO$ function. Note that darker colours of green and orange correspond to larger availability/needs of manure at some farm respectively. Figure \ref{fig:AIAappliedCat} is an indication that AIA solves the problem effectively.

\section{DISCUSSION}
\label{Discussion}
The work in this paper has addressed all the assumptions made in related work (see Section \ref{relWOrk}), being more detailed and complete Some assumptions have been addressed completely (assumptions 1, 2, 4, 6 and 7 as listed in Section \ref{relWOrk}) while some assumptions have been addressed partially (assumptions 3 and 5). For assumption 3, only Manhattan distances have been considered, while for assumption 5 the costs of purchasing/maintaining the vehicles used for the transfers have not been implemented. 

The AIA solution is completely decentralized, and could be well applied for a dynamic scenario in which animal manure production in livestock farms, as well as the actual needs in fertilizer in crop fields change over time. In the case of Catalonia, AIA transferred 51 tons of nitrogen, which constitute 62\% of the total needs in nitrogen of crop fields and 43\% of the total yearly availability of manure produced by animals in livestock farms. Figure \ref{fig:AIAappliedCat} indicates that further transfers of manure are not possible, as they do not yield positive financial outcomes. We suggest that the rest 57\% of the yearly manure production (ca. 66 tons) should be treated via local manure processing units, with possible conversion to bioenergy.

This study constitutes a demonstration that AIA could be employed for addressing this important problem.
A complete Life-Cycle Analysis (LCA) would consider a more comprehensive coverage of the problem, taking into account the extra costs needed to buy and maintain the vehicles used for the transfers (i.e. to compensate for the extra kilometres), as well as the extra time wasted by the livestock farmers or the personnel in charge of realizing the transfers of animal manure.

Future work will continue to explore the application of AIA to this problem, implementing more realistic transportation distances and travel times among farms for manure transport, as well as dynamic changes in production and need for nitrogen through the year. The possibility of using local manure processing units for manure, especially in larger livestock farms, will also be studied, under various policies that could be applied.

\section{CONCLUSIONS}
\label{Conclusion}
This paper addressed the important problem of the environmental impact of animal manure from livestock agriculture, 
considering a sustainable approach based on nutrient redistribution, where manure was transported as fertilizer from livestock farms to crop fields. A decentralized, dynamic approach was implemented, inspired by ant foraging behaviour (AIA), where the ants deposit pheromones near food sources in order to attract more ants to follow their trajectory. AIA addressed the problem by modelling livestock farms as ants and crop fields as sources of food for the ants. Results show that this approach is a promising solution to the problem, while the algorithm  works well and it is well balanced for the farmers in terms of transportation distances that need to be covered.

\section*{ACKNOWLEDGEMENTS}
Andreas Kamilaris has received funding from the European Union’s Horizon 2020 research and innovation programme under grant agreement No 739578 complemented by the Government of the Republic of Cyprus through the Directorate General for European Programmes, Coordination and Development.

{
	\begin{spacing}{1.17}
		\normalsize
		\bibliography{acobib} 
	\end{spacing}
}


\end{document}